\documentclass[aip,reprint,author-year,nofootinbib]{revtex4-1}

\usepackage{graphicx}
\usepackage{amsmath,amssymb}             
\usepackage{stmaryrd}					
\usepackage{color}
\usepackage[dvipsnames]{xcolor}
\usepackage{color,hyperref}
\definecolor{darkblue}{rgb}{0.0,0.0,0.4}
\definecolor{red}{rgb}{0.7,0.0,0.0}
\definecolor{green}{rgb}{0.0,0.5,0.0}
\hypersetup{colorlinks,breaklinks,
            linkcolor=darkblue,urlcolor=darkblue,
            anchorcolor=darkblue,citecolor=RoyalBlue}

\def\apj{Astrophys. J.}

\def\mnras{Mon. Not. R. Astron. Soc.}

\def\prd{Phys. Rev. D}

\def\physrep{Physics Reports}

\def\aap{Astron. \& Astrophys.}

\def\jcap{Journal of Cosmology and Astroparticle Physics}

\defcitealias{Bernardeau2002}{BCGS02}
\defcitealias{Weinberg1992}{W92}

\begin{document}


\title{One-point statistics of the Lagrangian displacement field}


\author{Florent Leclercq}
\email{florent.leclercq@polytechnique.org}
\affiliation{Institut d'Astrophysique de Paris (IAP), UMR 7095, CNRS -- UPMC Universit\'e Paris 6, Sorbonne Universit\'es, 98bis boulevard Arago, F-75014 Paris, France}
\affiliation{Institut Lagrange de Paris (ILP), Sorbonne Universit\'es,\\ 98bis boulevard Arago, F-75014 Paris, France}
\affiliation{\'Ecole polytechnique ParisTech,\\ Route de Saclay, F-91128 Palaiseau, France}

\author{Jens Jasche}
\affiliation{Excellence Cluster Universe, Technische Universit\"at M\"unchen,\\ Boltzmannstrasse 2, D-85748 Garching, Germany}

\author{Benjamin Wandelt}
\affiliation{Institut d'Astrophysique de Paris (IAP), UMR 7095, CNRS -- UPMC Universit\'e Paris 6, Sorbonne Universit\'es, 98bis boulevard Arago, F-75014 Paris, France}
\affiliation{Institut Lagrange de Paris (ILP), Sorbonne Universit\'es,\\ 98bis boulevard Arago, F-75014 Paris, France}
\affiliation{Department of Physics, University of Illinois at Urbana-Champaign,\\ 1110 West Green Street, Urbana, IL~61801, USA}
\affiliation{Department of Astronomy, University of Illinois at Urbana-Champaign,\\ 1002 West Green Street, Urbana, IL~61801, USA}


\date{\today}

\begin{abstract}
\noindent This document is an addendum to ``One-point remapping of Lagrangian perturbation theory in the mildly non-linear regime of cosmic structure formation'' \citep{Leclercq2013}.
\end{abstract}


\maketitle



The remapping procedure, described in section II of the main paper \citep{Leclercq2013}, relies on the Eulerian density contrast. As noted by previous authors \citep[see in particular][]{Neyrinck2013a}, in the Lagrangian representation of the large-scale structure, it is natural to use the divergence of the displacement field $\psi$ instead of the Eulerian density contrast $\delta$. This addendum provides additional comments on the one-point statistics of $\psi$ and comparatively analyzes key features of $\psi$ and $\delta$.

In the Lagrangian frame, the quantity of interest is not the position, but the displacement field $\Psi(\textbf{q})$ which maps the initial comoving particle position $\textbf{q}$ to its final comoving Eulerian position \textbf{x} (see e.g. \citealp{Bouchet1995} or \citealp{Bernardeau2002} for overviews),
\begin{equation}
\label{eq:Lagrangian-Eulerian-mapping}
\textbf{x} \equiv \textbf{q} + \Psi(\textbf{q}) .
\end{equation}
It is important to note that, though $\Psi(\textbf{q})$ is \emph{a priori} a full three-dimensional vector field, it is curl-free up to second order in Lagrangian perturbation theory (appendix D in \citealp{Bernardeau1994} or \citealp{Bernardeau2002} for a review). We did not consider contributions beyond 2LPT. After publication of the main paper, \citet{Chan2014} analyzed the non-linear evolution of $\Psi$, splitting it into its scalar and vector parts (the so-called ``Helmholtz decomposition''). Looking at two-point statistics, he found that shell-crossing leads to a suppression of small-scale power in the scalar part, and, subdominantly, to the generation of a vector contribution.

Let $\psi(\textbf{q}) \equiv \nabla_{\textbf{q}} \cdot \Psi(\textbf{q})$ denote the divergence of the displacement field, where $\nabla_{\textbf{q}}$ is the divergence operator in Lagrangian coordinates. $\psi$ quantifies the angle-averaged spatial-stretching of the Lagrangian dark matter ``sheet'' in comoving coordinates \citep{Neyrinck2013a}. Let $\mathcal{P}_{\psi,\mathrm{LPT}}$ and $\mathcal{P}_{\psi,\mathrm{Nbody}}$ be the one-point probability distribution functions for the divergence of the displacement field in LPT and in $N$-body fields, respectively. We denote by $\mathcal{P}_\delta$ the corresponding PDFs for the Eulerian density contrast.

\begin{figure}
\begin{center}
\includegraphics[width=\columnwidth]{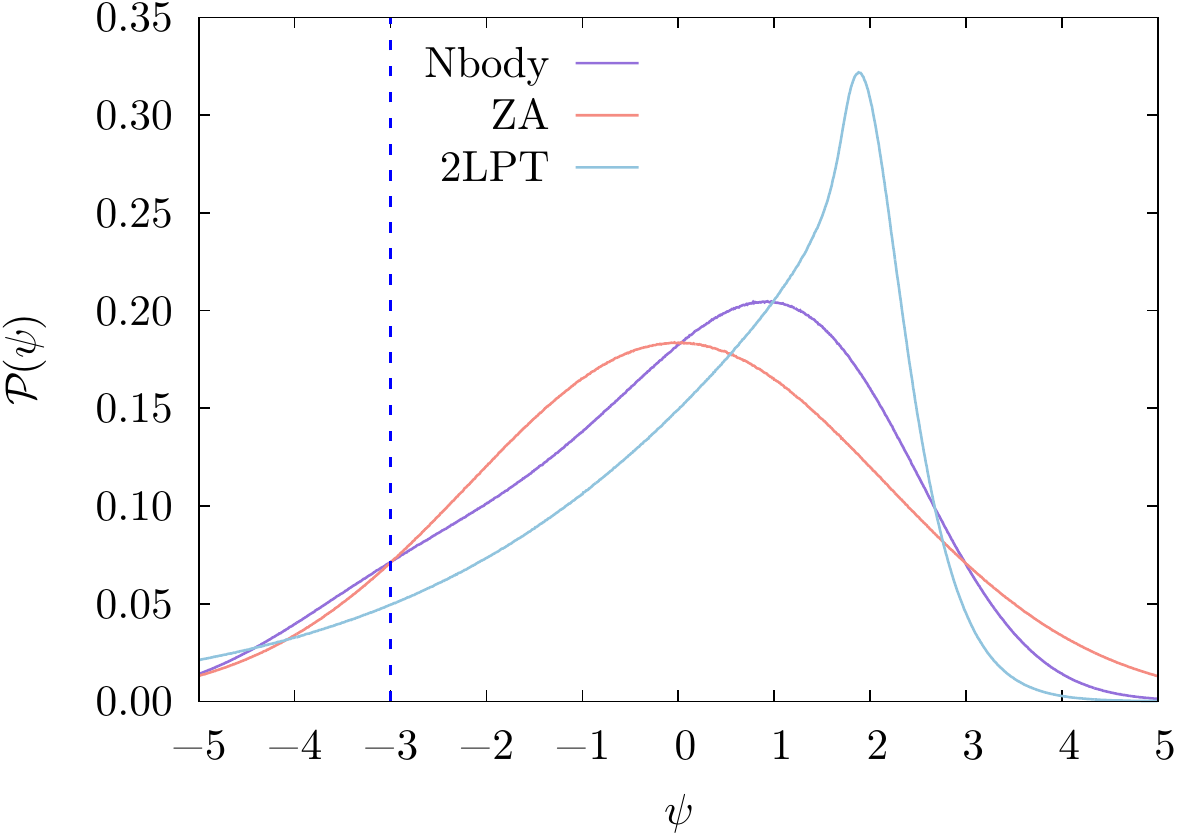}
\caption{Redshift-zero probability distribution function for the divergence of the displacement field $\psi$, computed from eight 1024~Mpc/$h$-box simulations of $512^3$ particles. A quantitative analysis of the deviation from Gaussianity of these PDFs is given in table \ref{tb:NGparam}. The particle distribution is determined using: a full $N$-body simulation (purple curve), the Zel'dovich approximation (ZA, light red curve) and second-order Lagrangian perturbation theory (2LPT, light blue curve). The vertical line at $\psi =-3$ represents the collapse barrier about which $\psi$ values bob around after gravitational collapse. A bump at this value is visible with full gravity, but LPT is unable to reproduce this feature. This regime corresponds to virialized, overdense clusters.}
\label{fig:divpsi_distrib}
\end{center}
\end{figure}

\begin{table}\centering
\begin{tabular}{lcc}
\hline\hline
Model & $\mathcal{P}_\delta$ & $\mathcal{P}_\psi$ \\
\hline
\multicolumn{1}{c}{} & \multicolumn{2}{c}{Skewness $\gamma_1$} \\
ZA & $2.36 \pm 0.01$ & $-0.0067 \pm 0.0001$ \\
2LPT & $2.83 \pm 0.01$ & $-1.5750 \pm 0.0002$ \\
$N$-body & $5.14 \pm 0.05$ & $-0.4274 \pm 0.0001$ \\
\hline
\multicolumn{1}{c}{} & \multicolumn{2}{c}{Excess kurtosis $\gamma_2$} \\
ZA & $9.95 \pm 0.09$ & $-2.2154 \times 10^{-6} \pm 0.0003$ \\
2LPT & $13.91 \pm 0.15$ & $3.544 \pm 0.0011$ \\
$N$-body & $62.60 \pm 2.75$ & $-0.2778 \pm 0.0004$ \\
\hline\hline
\end{tabular}
\caption{Non-Gaussianity parameters (the skewness $\gamma_1$ and the excess kurtosis $\gamma_2$) of the redshift-zero probability distribution functions $\mathcal{P}_\delta$ and $\mathcal{P}_\psi$ of the density contrast $\delta$ and the divergence of the displacement field $\psi$, respectively. The confidence intervals given correspond to the 1-$\sigma$ standard deviations among eight realizations. In all cases, $\gamma_1$ and $\gamma_2$ are reduced when measured from $\psi$ instead of $\delta$.}
\label{tb:NGparam}
\end{table}

In figure \ref{fig:divpsi_distrib}, we show the PDFs of $\psi$ for the ZA, 2LPT and full $N$-body gravity. The most important feature of $\psi$ is that, whatever the model for structure formation, the PDF exhibits reduced non-Gaussianity compared to the PDF for the density contrast $\delta$ \citep[see the upper panel of figure 7 in][for comparison]{Leclercq2013}. The main reason is that $\mathcal{P}_{\delta}$, unlike $\mathcal{P}_\psi$, is tied down to zero at $\delta = -1$. It is highly non-Gaussian in the final conditions, both in $N$-body simulations and in approximations to the true dynamics. For a quantitative analysis, we looked at the first and second-order non-Gaussianity statistics: the skewness $\gamma_1$ and the excess kurtosis $\gamma_2$,
\begin{equation}
\gamma_1 \equiv \frac{\mu_3}{\sigma^3} \quad \mathrm{and} \quad \gamma_2 \equiv \frac{\mu_4}{\sigma^4}-3,
\end{equation}
where $\mu_n$ is the $n$-th moment about the mean and $\sigma$ is the standard deviation. We estimated $\gamma_1$ and $\gamma_2$ at redshift zero in our simulations, in the one-point statistics of the density contrast $\delta$ and of the divergence of the displacement field $\psi$. The results are shown in table \ref{tb:NGparam}. In all cases, we found that both $\gamma_1$ and $\gamma_2$ are much smaller when measured from $\mathcal{P}_\psi$ instead of $\mathcal{P}_\delta$.

At linear order in Lagrangian perturbation theory (the Zel'dovich approximation), the divergence of the displacement field is proportional to the density contrast in the initial conditions, $\delta(\mathbf{q})$, scaling with the negative growth factor, $-D_1(\tau)$:
\begin{equation}
\label{eq:mapping-ZA}
\psi^{(1)}(\textbf{q},\tau) = \nabla_\textbf{q} \cdot \Psi^{(1)}(\textbf{q},\tau) = -D_1(\tau) \delta(\textbf{q}) .
\end{equation}
Since we take Gaussian initial conditions, the PDF for $\psi$ is Gaussian at any time with the ZA. In full gravity, non-linear evolution slightly breaks Gaussianity. $\mathcal{P}_{\psi,\mathrm{Nbody}}$ is slightly skewed towards negative values while its mode gets shifted around $\psi \approx 1$. Taking into account non-local effects, 2LPT tries to get closer to the shape observed in $N$-body simulations, but the correct skewness is overshot and the PDF is exceedingly peaked.

\begin{figure*}
\begin{center}
\includegraphics[width=0.85\textwidth]{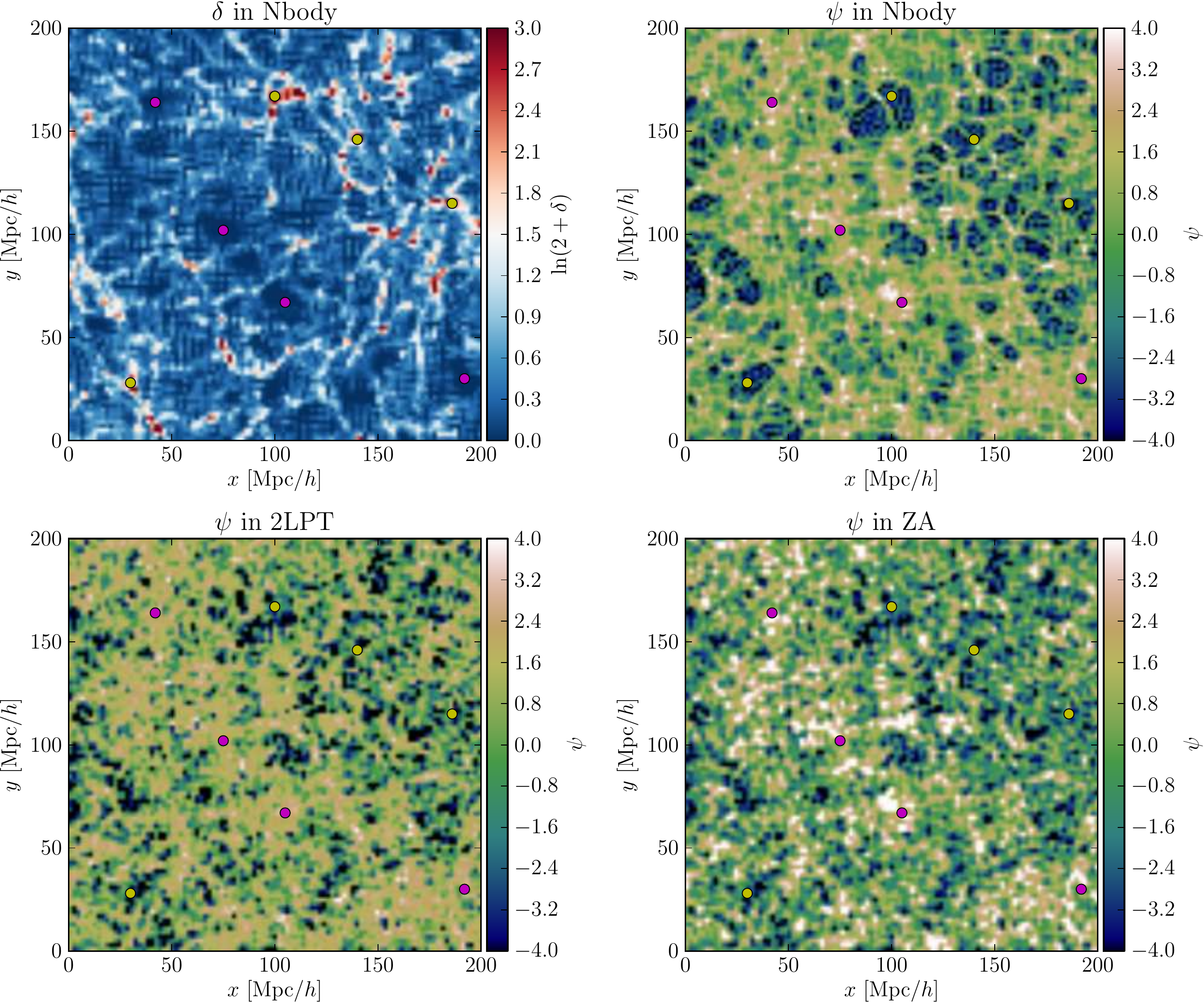}
\caption{Slices of the divergence of the displacement field, $\psi$, on a Lagrangian sheet of $512^2$ particles from a $512^3$-particle simulation of box size 1024 Mpc/$h$, run to redshift zero. For clarity we show only a 200~Mpc/$h$ region. Each pixel corresponds to a particle. The particle distribution is determined using respectively a full $N$-body simulation, the Zel'dovich approximation (ZA) and second-order Lagrangian perturbation theory (2LPT). In the upper left panel, the density contrast $\delta$ in the $N$-body simulation is shown, after binning on a $512^3$-voxel grid. To guide the eye, some clusters and voids are identified by yellow and purple dots, respectively. The ``lakes'', Lagrangian regions that have collapsed to form halos, are only visible in the $N$-body simulation, while the ``mountains'', Lagrangian regions corresponding to cosmic voids, are well reproduced by LPT.}
\label{fig:slices_divpsi}
\end{center}
\end{figure*}

\begin{figure*}
\begin{center}
\includegraphics[width=0.35\textwidth]{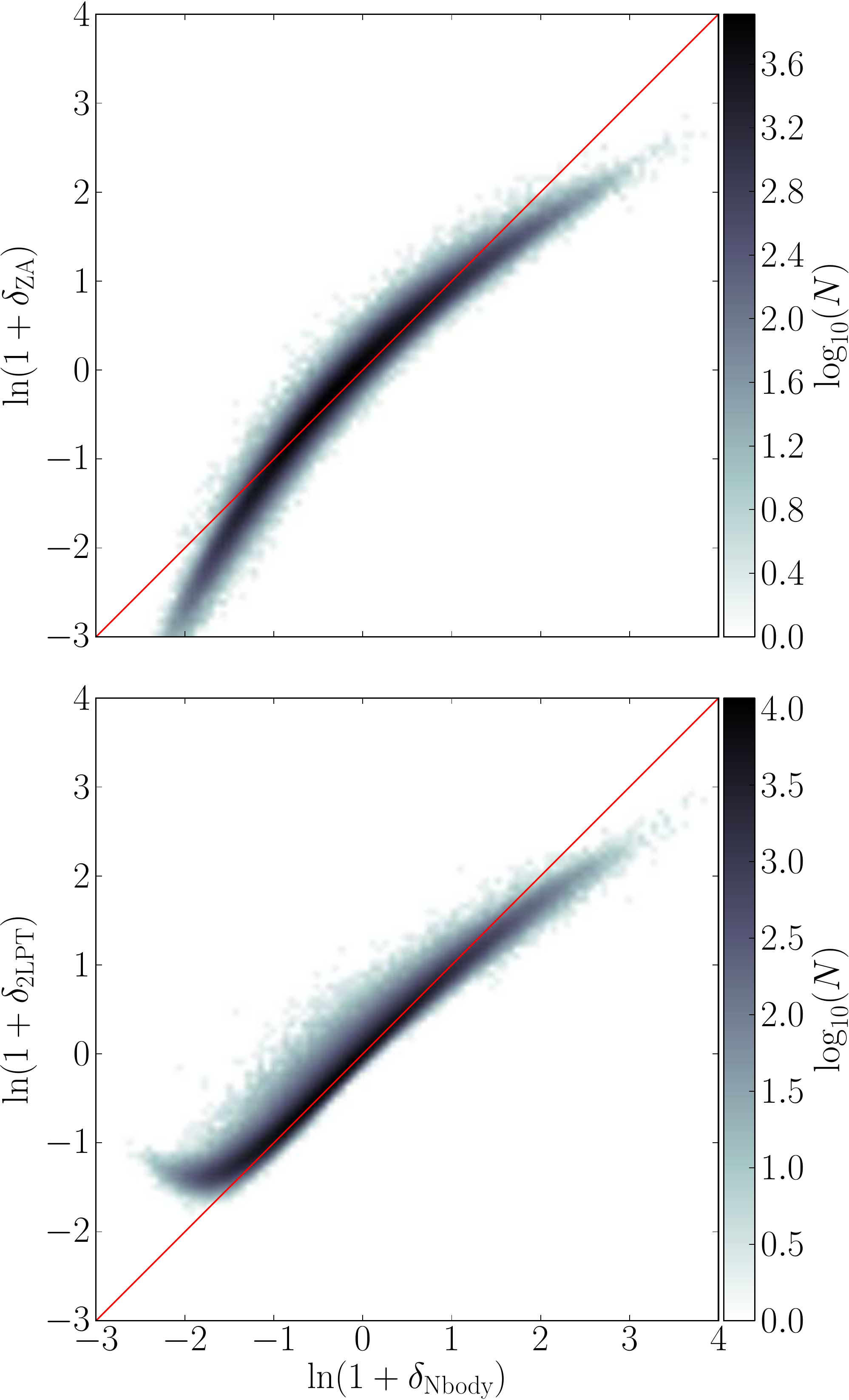} \quad \includegraphics[width=0.35\textwidth]{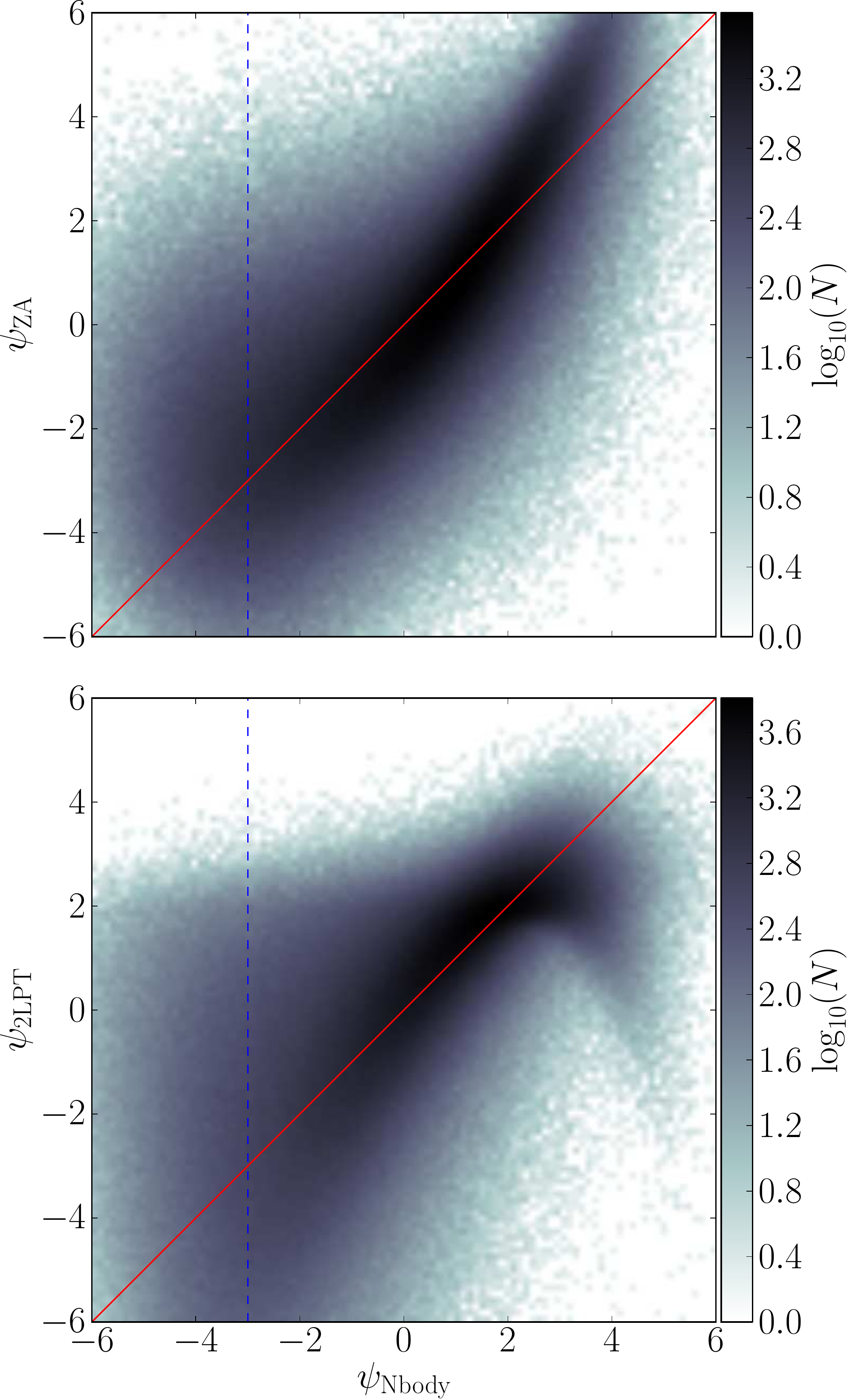}
\caption{\emph{Left panel}. Two-dimensional histograms comparing particle densities evolved with full $N$-body dynamics (the $x$-axis) to densities in the LPT-evolved particle distributions (the $y$-axis). The red lines show the ideal $y = x$ locus. A turn-up at low densities is visible with 2LPT, meaning that some overdense regions are predicted where there should be deep voids. \emph{Right panel}. Same plot for the divergence of the displacement field $\psi$. Negative $\psi$ corresponds to overdensities and positive $\psi$ correspond to underdensities. The dotted blue line shows the collapse barrier at $\psi = -3$ where particle get clustered in full gravity. The scatter is bigger with $\psi$ than with $\delta$, in particular in overdensities, since with LPT, particles do not cluster. The turn-up at low densities with 2LPT, observed with the density contrast, is also visible with the divergence of the displacement field.}
\label{fig:scatters}
\end{center}
\end{figure*}

Figure \ref{fig:slices_divpsi} shows a slice of the divergence of the displacement field, measured at redshift zero for particles occupying a flat $512^2$-pixel Lagrangian sheet from one of our simulations. For comparison, see also the figures in \citet{Mohayaee2006,Pueblas2009,Neyrinck2013a}. We used the color scheme of the latter paper, suggesting a topographical analogy when working in Lagrangian coordinates. As structures take shape, $\psi$ departs from its initial value; it takes positive values in underdensities and negative values in overdensities. The shape of voids (the ``mountains'') is found to be reasonably similar in LPT and in the $N$-body simulation. For this reason, the influence of late-time non-linear effects in voids is milder as compared to overdense structures, which makes them easier to relate to the initial conditions. However, in overdense regions where $\psi$ decreases, it is not allowed to take arbitrary values: where gravitational collapse occurs, ``lakes'' form and $\psi$ gets stuck around a collapse barrier, $\psi \approx -3$. As expected, these ``lakes'', corresponding to virialized clusters, can only be found in $N$-body simulations, since LPT fails to accurately describe the highly non-linear physics involved. A small bump at $\psi=-3$ is visible in $\mathcal{P}_{\psi,\mathrm{Nbody}}$ (see figure \ref{fig:divpsi_distrib}). We checked that this bump gets more visible in higher mass-resolution simulations (200 Mpc/$h$ box for $256^3$ particles), where matter is more clustered. This means that part of the information about gravitational clustering can be found in the one-point statistics of $\psi$. Of course, the complete description of halos requires to precisely account for the shape of the ``lakes'', which can only be done via higher-order correlation functions. More generally, it is possible to use Lagrangian information in order to classify structures of the cosmic web. In particular, \textsc{diva} \citep{Lavaux2010} uses the shear of the displacement field and \textsc{origami} \citep{Falck2012} the number of phase-space folds. As pointed out by \citet{Falck2015}, while these techniques cannot be straightforwardly used for the analysis of galaxy surveys, where we lack Lagrangian information, recently proposed techniques for physical inference of the initial conditions \citep{Jasche2013,Jasche2015} should allow their use with observational data.

Figure \ref{fig:scatters} shows two-dimensional histograms comparing $N$-body simulations to the LPT realizations for the density contrast $\delta$ and the divergence of the displacement field $\psi$. At this point, it is useful to note that a good mapping exists in the case where the relation shown is monotonic and the scatter is narrow. As pointed out by \citet{Sahni1996} and \citet{Neyrinck2013a}, matter in the substructure of 2LPT-voids has incorrect statistical properties: there are overdense particles in the low density region of the 2LPT $\delta$-scatter plot. This degeneracy is also visible in the $\psi > 0$ region of the 2LPT $\psi$-scatter plot. On average, the scatter is bigger with $\psi$ than with $\delta$, in particular in overdensities ($\psi <0$), since with LPT, particles do not cluster: $\psi$ takes any value between 2 and $-3$ where it should remain around $-3$.

Summing up our discussions in this addendum, we analyzed the relative merits of the Lagrangian divergence of the displacement field $\psi$, and the Eulerian density contrast $\delta$ at the level of one-point statistics. The important differences are the following:

\begin{enumerate}
\item $\Psi$ being irrotational up to order two, its divergence $\psi$ contains nearly all information on the displacement field in one dimension, instead of three. The collapse barrier at $\psi=-3$ is visible in $\mathcal{P}_\psi$ for $N$-body simulations but not for LPT. A part of the information about non-linear gravitational clustering is therefore encoded in the one-point statistics of $\psi$.
\item $\psi$ exhibits much fewer gravitationally-induced non-Gaussian features than $\delta$ in the final conditions (figure \ref{fig:divpsi_distrib} and table \ref{tb:NGparam}).
\item However, the values of $\psi$ are more scattered than the values of $\delta$ with respect to the true dynamics (figure \ref{fig:scatters}), meaning that an unambiguous mapping is more difficult.
\end{enumerate}

\textit{Note added}. While this addendum was being refeered, the work of \citet{Neyrinck2015} appeared. It uses the spherical collapse prescription for $\psi$ while checking various scales for the initial density field. The result is a fast scheme for producing approximate particle realizations.

\bibliography{Remapping_v4}

\begin{thebibliography}{15}%
\makeatletter
\providecommand \@ifxundefined [1]{%
 \@ifx{#1\undefined}
}%
\providecommand \@ifnum [1]{%
 \ifnum #1\expandafter \@firstoftwo
 \else \expandafter \@secondoftwo
 \fi
}%
\providecommand \@ifx [1]{%
 \ifx #1\expandafter \@firstoftwo
 \else \expandafter \@secondoftwo
 \fi
}%
\providecommand \natexlab [1]{#1}%
\providecommand \enquote  [1]{``#1''}%
\providecommand \bibnamefont  [1]{#1}%
\providecommand \bibfnamefont [1]{#1}%
\providecommand \citenamefont [1]{#1}%
\providecommand \href@noop [0]{\@secondoftwo}%
\providecommand \href [0]{\begingroup \@sanitize@url \@href}%
\providecommand \@href[1]{\@@startlink{#1}\@@href}%
\providecommand \@@href[1]{\endgroup#1\@@endlink}%
\providecommand \@sanitize@url [0]{\catcode `\\12\catcode `\$12\catcode
  `\&12\catcode `\#12\catcode `\^12\catcode `\_12\catcode `\%12\relax}%
\providecommand \@@startlink[1]{}%
\providecommand \@@endlink[0]{}%
\newcommand{\PineGreen}[1]{\textcolor{PineGreen}{#1}}%
\providecommand \url  [0]{\begingroup\@sanitize@url \@url }%
\providecommand \@url [1]{\endgroup\@href {#1}{\urlprefix }}%
\providecommand \urlprefix  [0]{URL }%
\providecommand \Eprint [0]{\href }%
\providecommand \doibase [0]{http://dx.doi.org/}%
\providecommand \selectlanguage [0]{\@gobble}%
\providecommand \bibinfo  [0]{\@secondoftwo}%
\providecommand \bibfield  [0]{\@secondoftwo}%
\providecommand \translation [1]{[#1]}%
\providecommand \BibitemOpen [0]{}%
\providecommand \bibitemStop [0]{}%
\providecommand \bibitemNoStop [0]{.\EOS\space}%
\providecommand \EOS [0]{\spacefactor3000\relax}%
\providecommand \BibitemShut  [1]{\csname bibitem#1\endcsname}%
\let\auto@bib@innerbib\@empty
\bibitem [{{Bernardeau}(1994)\citenamefont {{Bernardeau}}}]{Bernardeau1994}%
{(\PineGreen{{Bernardeau}}, \PineGreen{1994})}  \BibitemOpen
  \bibfield  {author} {\bibinfo {author} {\bibfnamefont {F.}~\bibnamefont
  {{Bernardeau}}},\ }\emph {{The nonlinear evolution of rare events}},\ \href
  {\doibase 10.1086/174121} {\bibfield  {journal} {\bibinfo  {journal} {\apj}\
  }\textbf {\bibinfo {volume} {427}},\ \bibinfo {pages} {51} (\bibinfo {year}
  {1994})},\ \Eprint {http://arxiv.org/abs/astro-ph/9311066}
  {astro-ph/9311066}\BibitemShut {NoStop}%
\bibitem [{{Bernardeau} {\textit{et~al}}\mbox{.}(2002)\citenamefont
  {{Bernardeau}}, \citenamefont {{Colombi}}, \citenamefont {{Gazta{\~n}aga}},\
  \&\ \citenamefont {{Scoccimarro}}}]{Bernardeau2002}%
{(\PineGreen{{Bernardeau} {\textit{et~al}}\mbox{.}}, \PineGreen{2002})}
  \BibitemOpen
  \bibfield  {author} {\bibinfo {author} {\bibfnamefont {F.}~\bibnamefont
  {{Bernardeau}}}, \bibinfo {author} {\bibfnamefont {S.}~\bibnamefont
  {{Colombi}}}, \bibinfo {author} {\bibfnamefont {E.}~\bibnamefont
  {{Gazta{\~n}aga}}}, \bibinfo {author} {\bibfnamefont {R.}~\bibnamefont
  {{Scoccimarro}}},\ }\emph {{Large-scale structure of the Universe and
  cosmological perturbation theory}},\ \href {\doibase
  10.1016/S0370-1573(02)00135-7} {\bibfield  {journal} {\bibinfo  {journal}
  {\physrep}\ }\textbf {\bibinfo {volume} {367}},\ \bibinfo {pages} {1}
  (\bibinfo {year} {2002})},\ \Eprint {http://arxiv.org/abs/astro-ph/0112551}
  {astro-ph/0112551}\BibitemShut {NoStop}%
\bibitem [{{Bouchet} {\textit{et~al}}\mbox{.}(1995)\citenamefont {{Bouchet}},
  \citenamefont {{Colombi}}, \citenamefont {{Hivon}},\ \&\ \citenamefont
  {{Juszkiewicz}}}]{Bouchet1995}%
{(\PineGreen{{Bouchet} {\textit{et~al}}\mbox{.}}, \PineGreen{1995})}
  \BibitemOpen
  \bibfield  {author} {\bibinfo {author} {\bibfnamefont {F.~R.}\ \bibnamefont
  {{Bouchet}}}, \bibinfo {author} {\bibfnamefont {S.}~\bibnamefont
  {{Colombi}}}, \bibinfo {author} {\bibfnamefont {E.}~\bibnamefont {{Hivon}}},
  \bibinfo {author} {\bibfnamefont {R.}~\bibnamefont {{Juszkiewicz}}},\ }\emph
  {{Perturbative Lagrangian approach to gravitational instability}},\
  \href@noop {} {\bibfield  {journal} {\bibinfo  {journal} {\aap}\ }\textbf
  {\bibinfo {volume} {296}},\ \bibinfo {pages} {575} (\bibinfo {year}
  {1995})},\ \Eprint {http://arxiv.org/abs/astro-ph/9406013}
  {astro-ph/9406013}\BibitemShut {NoStop}%
\bibitem [{{Chan}(2014)\citenamefont {{Chan}}}]{Chan2014}%
{(\PineGreen{{Chan}}, \PineGreen{2014})}  \BibitemOpen
  \bibfield  {author} {\bibinfo {author} {\bibfnamefont {K.~C.}\ \bibnamefont
  {{Chan}}},\ }\emph {{Helmholtz decomposition of the Lagrangian
  displacement}},\ \href {\doibase 10.1103/PhysRevD.89.083515} {\bibfield
  {journal} {\bibinfo  {journal} {\prd}\ }\textbf {\bibinfo {volume} {89}},\
  \bibinfo {eid} {083515} (\bibinfo {year} {2014})},\ \Eprint
  {http://arxiv.org/abs/1309.2243} {arXiv:1309.2243}\BibitemShut {NoStop}%
\bibitem [{{Falck} \& {Neyrinck}(2015)\citenamefont {{Falck}}\ \&\
  \citenamefont {{Neyrinck}}}]{Falck2015}%
{(\PineGreen{{Falck} \& {Neyrinck}}, \PineGreen{2015})}  \BibitemOpen
  \bibfield  {author} {\bibinfo {author} {\bibfnamefont {B.}~\bibnamefont
  {{Falck}}}, \bibinfo {author} {\bibfnamefont {M.~C.}\ \bibnamefont
  {{Neyrinck}}},\ }\emph {{The persistent percolation of single-stream
  voids}},\ \href {\doibase 10.1093/mnras/stv879} {\bibfield  {journal}
  {\bibinfo  {journal} {\mnras}\ }\textbf {\bibinfo {volume} {450}},\ \bibinfo
  {pages} {3239} (\bibinfo {year} {2015})},\ \Eprint
  {http://arxiv.org/abs/1410.4751} {arXiv:1410.4751}\BibitemShut {NoStop}%
\bibitem [{{Falck}, {Neyrinck} \& {Szalay}(2012)\citenamefont {{Falck}},
  \citenamefont {{Neyrinck}},\ \&\ \citenamefont {{Szalay}}}]{Falck2012}%
{(\PineGreen{{Falck}, {Neyrinck} \& {Szalay}}, \PineGreen{2012})}  \BibitemOpen
  \bibfield  {author} {\bibinfo {author} {\bibfnamefont {B.~L.}\ \bibnamefont
  {{Falck}}}, \bibinfo {author} {\bibfnamefont {M.~C.}\ \bibnamefont
  {{Neyrinck}}}, \bibinfo {author} {\bibfnamefont {A.~S.}\ \bibnamefont
  {{Szalay}}},\ }\emph {{ORIGAMI: Delineating Halos Using Phase-space Folds}},\
  \href {\doibase 10.1088/0004-637X/754/2/126} {\bibfield  {journal} {\bibinfo
  {journal} {\apj}\ }\textbf {\bibinfo {volume} {754}},\ \bibinfo {eid} {126}
  (\bibinfo {year} {2012})},\ \Eprint {http://arxiv.org/abs/1201.2353}
  {arXiv:1201.2353 [astro-ph.CO]}\BibitemShut {NoStop}%
\bibitem [{{Jasche}, {Leclercq} \& {Wandelt}(2015)\citenamefont {{Jasche}},
  \citenamefont {{Leclercq}},\ \&\ \citenamefont {{Wandelt}}}]{Jasche2015}%
{(\PineGreen{{Jasche}, {Leclercq} \& {Wandelt}}, \PineGreen{2015})}
  \BibitemOpen
  \bibfield  {author} {\bibinfo {author} {\bibfnamefont {J.}~\bibnamefont
  {{Jasche}}}, \bibinfo {author} {\bibfnamefont {F.}~\bibnamefont
  {{Leclercq}}}, \bibinfo {author} {\bibfnamefont {B.~D.}\ \bibnamefont
  {{Wandelt}}},\ }\emph {{Past and present cosmic structure in the SDSS DR7
  main sample}},\ \href {\doibase 10.1088/1475-7516/2015/01/036} {\bibfield
  {journal} {\bibinfo  {journal} {\jcap}\ }\textbf {\bibinfo {volume} {1}},\
  \bibinfo {eid} {036} (\bibinfo {year} {2015})},\ \Eprint
  {http://arxiv.org/abs/1409.6308} {arXiv:1409.6308}\BibitemShut {NoStop}%
\bibitem [{{Jasche} \& {Wandelt}(2013)\citenamefont {{Jasche}}\ \&\
  \citenamefont {{Wandelt}}}]{Jasche2013}%
{(\PineGreen{{Jasche} \& {Wandelt}}, \PineGreen{2013})}  \BibitemOpen
  \bibfield  {author} {\bibinfo {author} {\bibfnamefont {J.}~\bibnamefont
  {{Jasche}}}, \bibinfo {author} {\bibfnamefont {B.~D.}\ \bibnamefont
  {{Wandelt}}},\ }\emph {{Bayesian physical reconstruction of initial
  conditions from large-scale structure surveys}},\ \href {\doibase
  10.1093/mnras/stt449} {\bibfield  {journal} {\bibinfo  {journal} {\mnras}\
  }\textbf {\bibinfo {volume} {432}},\ \bibinfo {pages} {894} (\bibinfo {year}
  {2013})},\ \Eprint {http://arxiv.org/abs/1203.3639} {arXiv:1203.3639
  [astro-ph.CO]}\BibitemShut {NoStop}%
\bibitem [{{Lavaux} \& {Wandelt}(2010)\citenamefont {{Lavaux}}\ \&\
  \citenamefont {{Wandelt}}}]{Lavaux2010}%
{(\PineGreen{{Lavaux} \& {Wandelt}}, \PineGreen{2010})}  \BibitemOpen
  \bibfield  {author} {\bibinfo {author} {\bibfnamefont {G.}~\bibnamefont
  {{Lavaux}}}, \bibinfo {author} {\bibfnamefont {B.~D.}\ \bibnamefont
  {{Wandelt}}},\ }\emph {{Precision cosmology with voids: definition, methods,
  dynamics}},\ \href {\doibase 10.1111/j.1365-2966.2010.16197.x} {\bibfield
  {journal} {\bibinfo  {journal} {\mnras}\ }\textbf {\bibinfo {volume} {403}},\
  \bibinfo {pages} {1392} (\bibinfo {year} {2010})},\ \Eprint
  {http://arxiv.org/abs/0906.4101} {arXiv:0906.4101 [astro-ph.CO]}\BibitemShut
  {NoStop}%
\bibitem [{{Leclercq} {\textit{et~al}}\mbox{.}(2013)\citenamefont {{Leclercq}},
  \citenamefont {{Jasche}}, \citenamefont {{Gil-Mar{\'{\i}}n}},\ \&\
  \citenamefont {{Wandelt}}}]{Leclercq2013}%
{(\PineGreen{{Leclercq} {\textit{et~al}}\mbox{.}}, \PineGreen{2013})}
  \BibitemOpen
  \bibfield  {author} {\bibinfo {author} {\bibfnamefont {F.}~\bibnamefont
  {{Leclercq}}}, \bibinfo {author} {\bibfnamefont {J.}~\bibnamefont
  {{Jasche}}}, \bibinfo {author} {\bibfnamefont {H.}~\bibnamefont
  {{Gil-Mar{\'{\i}}n}}}, \bibinfo {author} {\bibfnamefont {B.}~\bibnamefont
  {{Wandelt}}},\ }\emph {{One-point remapping of Lagrangian perturbation theory
  in the mildly non-linear regime of cosmic structure formation}},\ \href
  {\doibase 10.1088/1475-7516/2013/11/048} {\bibfield  {journal} {\bibinfo
  {journal} {\jcap}\ }\textbf {\bibinfo {volume} {11}},\ \bibinfo {eid} {048}
  (\bibinfo {year} {2013})},\ \Eprint {http://arxiv.org/abs/1305.4642}
  {arXiv:1305.4642 [astro-ph.CO]}\BibitemShut {NoStop}%
\bibitem [{{Mohayaee} {\textit{et~al}}\mbox{.}(2006)\citenamefont {{Mohayaee}},
  \citenamefont {{Mathis}}, \citenamefont {{Colombi}},\ \&\ \citenamefont
  {{Silk}}}]{Mohayaee2006}%
{(\PineGreen{{Mohayaee} {\textit{et~al}}\mbox{.}}, \PineGreen{2006})}
  \BibitemOpen
  \bibfield  {author} {\bibinfo {author} {\bibfnamefont {R.}~\bibnamefont
  {{Mohayaee}}}, \bibinfo {author} {\bibfnamefont {H.}~\bibnamefont
  {{Mathis}}}, \bibinfo {author} {\bibfnamefont {S.}~\bibnamefont {{Colombi}}},
  \bibinfo {author} {\bibfnamefont {J.}~\bibnamefont {{Silk}}},\ }\emph
  {{Reconstruction of primordial density fields}},\ \href {\doibase
  10.1111/j.1365-2966.2005.09774.x} {\bibfield  {journal} {\bibinfo  {journal}
  {\mnras}\ }\textbf {\bibinfo {volume} {365}},\ \bibinfo {pages} {939}
  (\bibinfo {year} {2006})},\ \Eprint {http://arxiv.org/abs/astro-ph/0501217}
  {astro-ph/0501217}\BibitemShut {NoStop}%
\bibitem [{{Neyrinck}(2013)\citenamefont {{Neyrinck}}}]{Neyrinck2013a}%
{(\PineGreen{{Neyrinck}}, \PineGreen{2013})}  \BibitemOpen
  \bibfield  {author} {\bibinfo {author} {\bibfnamefont {M.~C.}\ \bibnamefont
  {{Neyrinck}}},\ }\emph {{Quantifying distortions of the Lagrangian
  dark-matter mesh in cosmology}},\ \href {\doibase 10.1093/mnras/sts027}
  {\bibfield  {journal} {\bibinfo  {journal} {\mnras}\ }\textbf {\bibinfo
  {volume} {428}},\ \bibinfo {pages} {141} (\bibinfo {year} {2013})},\ \Eprint
  {http://arxiv.org/abs/1204.1326} {arXiv:1204.1326 [astro-ph.CO]}\BibitemShut
  {NoStop}%
\bibitem [{{Neyrinck}(2015)\citenamefont {{Neyrinck}}}]{Neyrinck2015}%
{(\PineGreen{{Neyrinck}}, \PineGreen{2015})}  \BibitemOpen
  \bibfield  {author} {\bibinfo {author} {\bibfnamefont {M.~C.}\ \bibnamefont
  {{Neyrinck}}},\ }\emph {{Truthing the stretch: Non-perturbative cosmological
  realizations with multiscale spherical collapse}},\ \href@noop {} {\bibfield
  {journal} {\bibinfo  {journal} {ArXiv e-prints}\ } (\bibinfo {year}
  {2015})},\ \Eprint {http://arxiv.org/abs/1503.07534}
  {arXiv:1503.07534}\BibitemShut {NoStop}%
\bibitem [{{Pueblas} \& {Scoccimarro}(2009)\citenamefont {{Pueblas}}\ \&\
  \citenamefont {{Scoccimarro}}}]{Pueblas2009}%
{(\PineGreen{{Pueblas} \& {Scoccimarro}}, \PineGreen{2009})}  \BibitemOpen
  \bibfield  {author} {\bibinfo {author} {\bibfnamefont {S.}~\bibnamefont
  {{Pueblas}}}, \bibinfo {author} {\bibfnamefont {R.}~\bibnamefont
  {{Scoccimarro}}},\ }\emph {{Generation of vorticity and velocity dispersion
  by orbit crossing}},\ \href {\doibase 10.1103/PhysRevD.80.043504} {\bibfield
  {journal} {\bibinfo  {journal} {\prd}\ }\textbf {\bibinfo {volume} {80}},\
  \bibinfo {eid} {043504} (\bibinfo {year} {2009})},\ \Eprint
  {http://arxiv.org/abs/0809.4606} {arXiv:0809.4606}\BibitemShut {NoStop}%
\bibitem [{{Sahni} \& {Shandarin}(1996)\citenamefont {{Sahni}}\ \&\
  \citenamefont {{Shandarin}}}]{Sahni1996}%
{(\PineGreen{{Sahni} \& {Shandarin}}, \PineGreen{1996})}  \BibitemOpen
  \bibfield  {author} {\bibinfo {author} {\bibfnamefont {V.}~\bibnamefont
  {{Sahni}}}, \bibinfo {author} {\bibfnamefont {S.}~\bibnamefont
  {{Shandarin}}},\ }\emph {{Accuracy of Lagrangian approximations in voids}},\
  \href@noop {} {\bibfield  {journal} {\bibinfo  {journal} {\mnras}\ }\textbf
  {\bibinfo {volume} {282}},\ \bibinfo {pages} {641} (\bibinfo {year}
  {1996})},\ \Eprint {http://arxiv.org/abs/astro-ph/9510142}
  {astro-ph/9510142}\BibitemShut {NoStop}%
\end{thebibliography}%

\end{document}